# Transformation optics that mimics the system outside a Schwarzschild black hole

**Huanyang Chen,** [1, a] **Rong-Xin Miao,** [2,3] **and Miao Li** [2,3, b]

[1]*School of Physical Science and Technology, Soochow University, Suzhou, Jiangsu 215006, China*
[2]*Kavli Institute for Theoretical Physics, Key Laboratory of Frontiers in Theoretical Physics, Institute of Theoretical Physics, Chinese Academy of Sciences, Beijing 100190, People's Republic of China*
[3]*Interdisciplinary Center for Theoretical Study, University of Science and Technology of China, Hefei, Anhui 230026, China*
[a]*kenyon@ust.hk*
*http://ihome.ust.hk/~kenyon*
[b]*mli@itp.ac.cn*

**Abstract:** We applied the transformation optics to mimic a black hole of Schwarzschild form. Similar properties of photon sphere were also found numerically for the metamaterial black hole. Several reduced versions of the black hole systems were proposed for easier implementations.

©2009 Optical Society of America

## References and links

1. U. Leonhardt, "Optical conformal mapping," Science **312,** 1777-1780 (2006).
2. J. B. Pendry, D. Schurig, and D. R. Smith, "Controlling electromagnetic fields," Science **312,** 1780-1782 (2006).
3. D. Schurig, J. J. Mock, B. J. Justice, S. A. Cummer, J. B. Pendry, A. F. Starr, and D. R. Smith, "Metamaterial electromagnetic cloak at microwave frequencies," Science **314,** 977-980 (2006).
4. R. Liu, C. Ji, J. J. Mock, J. Y. Chin, T. J. Cui, D. R. Smith, "Broadband ground-plane cloak," Science **323,** 366-369 (2009).
5. S. Tretyakov, P. Alitalo, O. Luukkonen, and C. Simovski, "Broadband electromagnetic cloaking of long cylindrical objects," Phys. Rev. Lett. **103,** 103905 (2009).
6. J. Valentine, J. Li, T. Zentgraf, G. Bartal, and X. Zhang, "An optical cloak made of dielectrics," Nature Mater. **8,** 568-571 (2009).
7. L. H. Gabrielli, J. Cardenas, C. B. Poitras, and M. Lipson, "Silicon nanostructure cloak operating at optical frequencie," Nature Photonics **3,** 461-463 (2009).
8. I. I. Smolyaninov, V. N. Smolyaninova, A. V. Kildishev, and V. M. Shalaev, "Anisotropic metamaterials emulated by tapered waveguides: Application to optical cloaking," Phys. Rev. Lett. **102,** 213901 (2009).
9. H. Y. Chen, B. Hou, S. Chen, X. Ao, W. Wen, and C. T. Chan, "Design and experimental realization of a broadband transformation media field rotator at microwave frequencies," Phys. Rev. Lett. **102,** 183903 (2009).
10. Z. L. Mei and T. J. Cui, "Experimental realization of a broadband bend structure using gradient index metamaterials," Opt. Express **17,** 18354-18363 (2009).
11. Y. G. Ma, C. K. Ong, T. Tyc, and U. Leonhardt, "An omnidirectional retroreflector based on the transmutation of dielectric singularities," Nature Mater. **8,** 639-642 (2009).
12. U. Leonhardt and T. G. Philbin, "General relativity in electrical engineering," New J. Phys. **8,** 247 (2006).
13. D. A. Genov, S. Zhang, and X. Zhang, "Mimicking celestial mechanics in metamaterials," Nature Phys. **5,** 687-692 (2009).
14. E. E. Narimanov and A. V. Kildishev, "Optical black hole: Broadband omnidirectional light absorber," Appl. Phys. Lett. **95,** 041106 (2009).
15. Q. Cheng and T. J. Cui, "An electromagnetic black hole made of metamaterials," arXiv: 0910.2159.
16. A. Greenleaf, M. Lassas, and G. Uhlmann, "Electromagnetic wormholes and virtual magnetic monopoles from metamaterials," Phys. Rev. Lett. **99,** 183901 (2007).
17. M. Li, R.-X. Miao, and Y. Pang, "Casimir energy, holographic dark energy and electromagnetic metamaterial mimicking de Sitter," arXiv: 0910.3375.
18. T. G. Mackay and A. Lakhtakia, "Towards a metamaterial simulation of a spinning cosmic string," arXiv: 0911.4163.

19. We have designed a perfectly matched layer (PML) in the far field outside the black hole to mimic such an open system during the simulations.
20. U. Leonhardt and T.G. Philbin, "Quantum optics of spatial transformation media," J. Opt. A: Pure Appl. Opt. **9,** S289 (2007).
21. S. W. Hawking, "Black Hole Explosions," Nature **248,** 30-31 (1974).

## 1. Introduction

Transformation optics [1, 2] is a versatile tool that can be extensively utilized in designing novel wave manipulation devices. With the aid of metamaterials, several kinds of transformation media have been implemented in laboratories, such as reduced (or carpet) cloaks for both microwave [3-5] and optical frequencies [6-8], field rotator [9], transformation media bending wave guide [10], and transmuted Eaton lens [11]. It is surprising that transformation optics can even mimic the cosmic phenomena. Leonhardt and Philbin proposed the "general relativity in electrical engineering" [12] which can be used to design metamaterials mimicking the cosmic optical properties. The corresponding transformation media are usually anisotropic and inhomogeneous materials. Genov *et al.* [13] suggested a further method to transmute the anisotropic materials into isotropic ones, which suggests a much easier way to implement. An isotropic "optical black hole" was then proposed and studied [13]. Narimanov and Kildishev [14] independently suggested an approach to a broadband absorber from Hamiltonian optics, which was also termed as an effective "optical black hole". The related materials are isotropic as well, which was later implemented by using nonresonant metamaterial units for microwave frequencies by Cheng and Cui [15]. However, both versions of the optical black hole were not solutions to the Einstein field equations [13, 14]. Hence it is interesting to study how to mimic a real black hole system by using transformation optics and their corresponding properties. In this letter, we will start from the "general relativity in electrical engineering" [12], and use the Schwarzschild metric of a real black hole system (which is a solution to the Einstein field equations) to propose the permittivity and permeability tensors of the related transformation media. With the material parameters obtained, we will simulate wave properties outside the black hole (the Schwarzschild radius), and we are trying to observe the phenomena of "photon sphere", which is an important clue for the black hole system. Before we go into the details, it should be noted that transformation optics has also been applied to other cosmic problems, such as electromagnetic wormhole [16], de Sitter space [17] and cosmic string [18].

## 2. Theory

Let us now start from the general form of transformation optics [12],

$$\varepsilon^{ij} = \mu^{ij} = \mp \frac{\sqrt{-g}}{\sqrt{\gamma}} \frac{g^{ij}}{g_{00}}, \qquad (1)$$

where the $\varepsilon^{ij}$ and $\mu^{ij}$ are the equivalent material parameters of the transformation media to mimic a space whose metric can be written in the form of $ds^2 = \sum_{\alpha\beta} g_{\alpha\beta} dx^\alpha dx^\beta$, $\gamma$ is a determinant of a spatial metric to transform one set of spatial coordinates to another [12]. The Schwarzschild metric of a black hole system is written as,

$$ds^2 = (1 - \frac{2GM}{rc^2})c^2 dt^2 - (1 - \frac{2GM}{rc^2})^{-1} dr^2 - r^2 d\Omega^2, \qquad (2)$$

where $G$ is the gravitational constant, $M$ is the mass of the system, $c$ is the velocity of light in vacuum. With respect to the optical phenomena in the equatorial plane, it is equivalent to consider the metric in circular cylindrical coordinates,

$$ds^2 = (1-\frac{L}{r})c^2 dt^2 - (1-\frac{L}{r})^{-1} dr^2 - r^2 d\theta^2 - dz^2, \qquad (3)$$

which has the same optical behaviors as that of a real Schwarzschild black hole, and is easier to realize in laboratory than the original three dimensional (3D) case.

After some algebraic calculations from Eqs. (1) and (3), we have,

$$\varepsilon^{ij} = \mu^{ij} = \frac{1}{1-\frac{L}{r}} \begin{pmatrix} 1-\frac{L}{r}\frac{x^2}{r^2} & -\frac{L}{r}\frac{xy}{r^2} & 0 \\ -\frac{L}{r}\frac{xy}{r^2} & 1-\frac{L}{r}\frac{y^2}{r^2} & 0 \\ 0 & 0 & 1 \end{pmatrix}, \qquad (4)$$

for the transformation media to mimic the black hole system, with its principal values of the permittivity and permeability tensors in circular cylindrical coordinates as,

$$\varepsilon_r = \mu_r = 1, \quad \varepsilon_\theta = \mu_\theta = \varepsilon_z = \mu_z = (1-\frac{L}{r})^{-1}. \qquad (5)$$

For a 3D black hole system, with similar calculations, the parameters should be,

$$\varepsilon_r = \mu_r = 1, \quad \varepsilon_\theta = \mu_\theta = \varepsilon_\varphi = \mu_\varphi = (1-\frac{L}{r})^{-1}, \qquad (6)$$

in spherical coordinates.

In this letter, we shall only consider the transverse electric (TE) polarized cases in two dimensions (2D) for simplicity. As the parameters diverge when $r$ approaches the event horizon $L$ and the problem cannot be studied analytically, we shall have to do some truncations to tackle the singularity so that the problem can be studied numerically. It would be wise to choose the following material parameters,

$$\mu_r = 1, \quad \mu_\theta = \varepsilon_z = (1-\frac{L_1}{r})^{-1}, \quad r > L, \qquad (7a)$$

$$\mu_r = 1, \quad \mu_\theta = \varepsilon_z = (1-\frac{L_1}{L})^{-1}(1+i), \quad 0 < r < L, \qquad (7b)$$

for the black hole system, where $L_1$ is a slightly smaller number than $L$ (here we set $L_1 = 0.9L$ so that the largest values of $\mu_\theta$ and $\varepsilon_z$ are 10 when $r = L$). All the light should be absorbed when they hit the event horizon ideally. Thereby we choose the above impedance-matched absorbing core [13] (Eq. (7b), $\mu_r = 1, \mu_\theta = \varepsilon_z = 10+10i$ ) to approximate the region inside the event horizon. Most of the light will be absorbed by the core when they reach the event horizon and little scatterings should be caused. It is noted that the material parameters in Eq. (7) can be implemented using the same method as that in the first TE reduced cloak [3]. The only difference is that, the TE cloak takes a $\mu_\theta = 1$ [3] while the black hole here uses a $\mu_r = 1$. Therefore during the implementation of the black hole, the normal directions of the split-ring resonators [3] should be along the angular directions.

### 3. Simulation results

For a real black hole system there is a photon sphere (here we shall call it the "photon cylinder" with respect to the circular cylindrical coordinate used in Eq. (7)) where the gravity is so strong to force the photons to orbit. The radius of the sphere (or cylinder) should be 1.5 times of the Schwarzschild radius. In Fig. 1 (a), we plot the emitting rays of a point source when it is on the photon sphere $(1.5L_1, 0)$ of the real black hole. The rays (green lines) will escape from the black hole if the radial component of the wavevector is positive and the rays (red lines) of wavevector with negative radial components cannot escape and hence be absorbed by the core. As a result the energy escaped should be equal to the energy absorbed. It is noted that interference is indicated by the coinciding rays escaping from both sides of the black hole. In Fig. 1 (b), we plot the electric fields when a line source is located on the photon cylinder near the black hole described by Eq. (7). Interference is clearly seen meaning a possibly "photon cylinder" is created as in the real black hole.

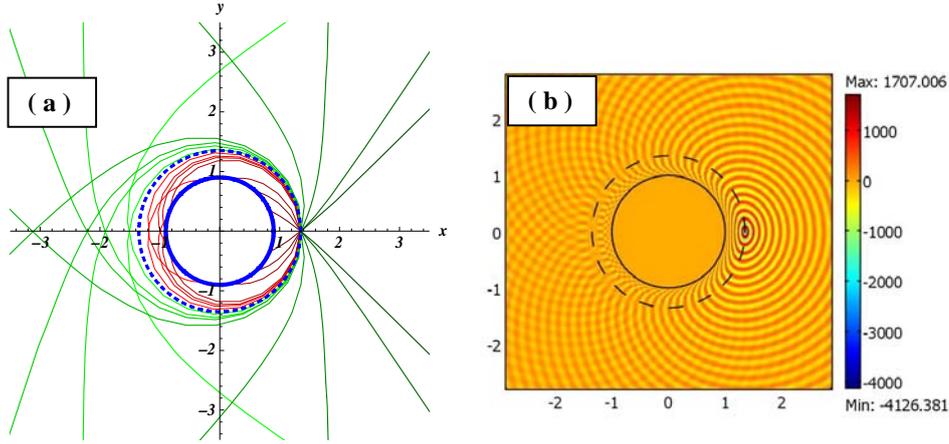

Fig. 1. (a) The emitting rays of a point source located on the photon sphere at $(1.5L_1, 0)$ for a real black hole. The blue dashed circle outlines the photon sphere while the blue solid circle is the event horizon. The green lines (outside the dashed circle) demonstrate the rays that can escape from the black hole while the red lines (inside the dashed circle and outside the solid circle) show the ones absorbed. (b) The electric fields outside the event horizon when a line source is located on the photon cylinder described by Eq. (7). The dashed circle outlines the photon cylinder while the solid circle is the event horizon. The incident wavelength is $\lambda = 0.25L$. The simulation was carried out by using the COMSOL Multiphysics finite element-based electromagnetics solver.

In Fig. 2, we demonstrate the simulations for different incident Gaussian beams (which are composed of different wavevectors with majority of them pointing in the propagation directions) interacting with the metamaterial black hole. Figure 2 (a) shows a Gaussian beam propagating close to the photon cylinder. Clearly the beam is bent enormously toward the core region as if there is an attracting force. Most of the wavevectors have negative radial components and thus will be absorbed by the core. If the beam is moved farther from the photon cylinder (see Fig. 2 (b)), we will reach an interesting situation that most wavevectors will be along the angular directions and orbit almost circularly around the core and in this way the photons will be trapped inside the orbit for a tremendous amount of time. In Fig. 2 (c), we move the beam even farther. Most of the wavevectors are still bent with almost comparable amount of them escaping and absorbed. Finally we plot in Fig. 2 (d) to show the situation of a much farther beam with most part of it escaping from the black hole. All these demonstrate a strong similarity with the photon sphere in the real black hole.

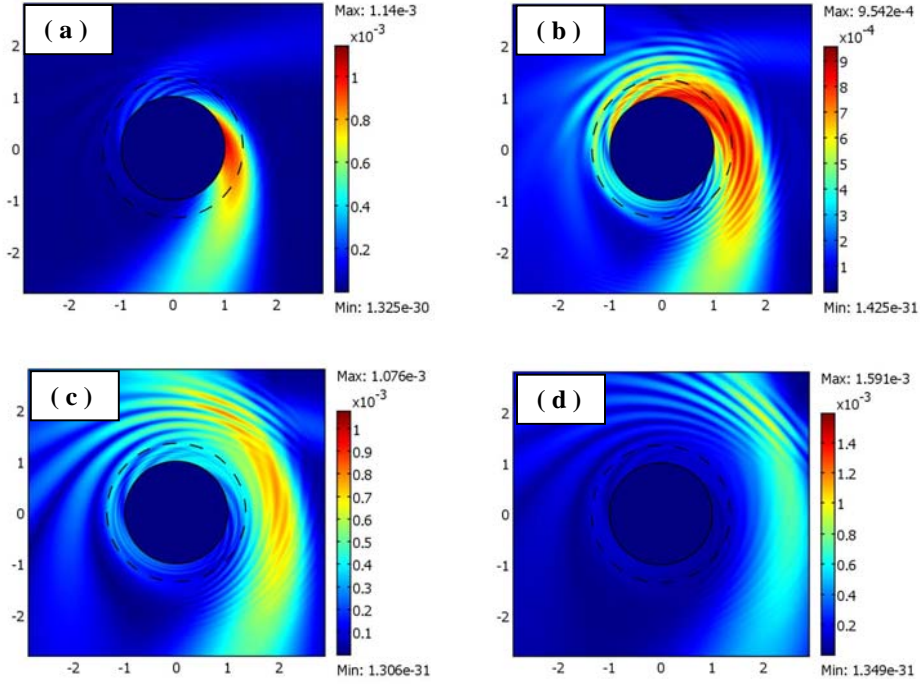

Fig. 2. The magnetic field amplitude $|\vec{H}|$ near the black hole from Eq. (7) for different incident Gaussian beams with beam centers at (a) $(2L, 0)$, (b) $(2.5L, 0)$, (c) $(3L, 0)$, (d) $(3.5L, 0)$. The incident wavelengths and the beam widths are $\lambda = w = 0.25L$. The beams are incident at an angle of 60 *deg* from x-axis. The simulation was carried out by using the COMSOL Multiphysics finite element-based electromagnetics solver.

## 4. Parameter reductions

It is still very challenging to implement such a metamaterial black hole system in two reasons. One is that the device is infinitely large [19] and filled in the whole space. The other is that it composes of anisotropic materials. It would be worth trying to use another metric,

$$ds^2 = \alpha(1-\frac{L}{r})c^2dt^2 - \alpha^{-1}(1-\frac{L}{r})^{-1}dr^2 - r^2d\theta^2 - dz^2, \qquad (8)$$

where we can choose the value of $\alpha$ to be $(1-L/L_2)^{-1}$ so that when $r = L_2$, $\alpha(1-L/r) = 1$. The material parameters of the transformation media inside $L \leq r \leq L_2$ should then be,

$$\varepsilon_r = \mu_r = 1, \quad \varepsilon_\theta = \mu_\theta = \varepsilon_z = \mu_z = (\alpha - \alpha\frac{L}{r})^{-1}, \qquad (9)$$

while outside $r = L_2$ it is vacuum, so that the device becomes finite for feasible implementation. Further transformation can be performed to transmute the transformation media described by Eq. (5) or (9) into isotropic materials following the methods given by Genov *et al.* [13]. For example, we can use the following transformation from $r$ to $\tilde{r}$,

$$r(\tilde{r}) = \frac{16\tilde{r}^2 + 8\tilde{r}L + L^2}{16\tilde{r}}, \qquad (10)$$

so that we can obtain an isotropic "black hole" from Eq. (5) with its refractive index profile as,

$$n(\tilde{r}) = \frac{(4\tilde{r} + L)^3}{16\tilde{r}^2(4\tilde{r} - L)}, \qquad (11)$$

for $\tilde{r} \in [\frac{L}{4}, \infty]$. Similar transmutation can be performed to Eq. (9). Although the transformation media of Eq. (9) or (11) are not exactly mimicking the real black hole systems that have solutions to Einstein field equations, they are still quite useful versions to demonstrate the above "photon sphere" properties in future proof-of-principle experiments. To implement the above reduced black hole, one has to truncate the parameters near the event horizon due to the singularities [13]. In addition, it should be also very interesting to see whether one can use the transformation media concept (such as the "quantum optics of spatial transformation media" [20]) to mimic the behaviors inside the black hole (the absorbing core) and explore the Hawking radiation [21] effect in the mimicked metamaterial systems.

**5. Conclusions**

To summarize, we have used the transformation media concept to mimic a system outside the black hole of Schwarzschild form. The similar "photon sphere" properties were found in simulation results. Several reduced versions of black holes were also proposed for future proof-of-principle experimental demonstrations. We showed transformation media concept have versatility in studying the cosmic phenomena.

**Acknowledgments**

This work was supported by the Soochow University Start-up grant No. Q4108909, the NSFC grant No. 10535060/A050207, a NSFC group grant No. 10821504 and Ministry of Science and Technology 973 program under grant No. 2007CB815401.